\documentclass[aps,onecolumn,superscriptaddress,nofootinbib]{revtex4}
\usepackage{graphicx}
\usepackage{dcolumn}
\usepackage{amssymb}
\usepackage{bm}
\usepackage{enumerate}
\begin{document}
\def\Universita{Universit\`a}
\title{\bf \LARGE Comments
 on ``Improved limit on
  quantum-spacetime modifications of Lorentz symmetry
  from observations of gamma-ray blazars"}
\author{T. Jacobson, S. Liberati, and D. Mattingly\\
{\small \it
Department of Physics University of Maryland, College Park, MD
20742-4111, USA} }

\begin{abstract}
We address several criticisms by  Amelino-Camelia of our recent
analyses of two observational constraints on Lorentz violation at
order $E/M_{\rm Planck}$. In particular, we emphasize the role of
effective field theory in our analysis of synchrotron radiation,
and we strengthen the justification for the constraint coming
from photon annihilation.
\end{abstract}

\maketitle
\def\wt{\widetilde}
\def\gsim{\; \raisebox{-.8ex}{$\stackrel{\textstyle >}{\sim}$}\;}
\def\lsim{\; \raisebox{-.8ex}{$\stackrel{\textstyle <}{\sim}$}\;}
\def\half{{1\over2}}
\def\a{\alpha}
\def\b{\beta}
\def\g{\gamma}
\def\e{\epsilon}
\def\m{\mu}
\def\o{\omega}
\def\L{{\mathcal L}}
\def\d{{\mathrm{d}}}
\def\p{{\mathbf{p}}}
\def\q{{\mathbf{q}}}
\def\k{{\mathbf{k}}}
\def\fp{{p_{\rm 4}}}
\def\fq{{q_{\rm 4}}}
\def\fk{{k_{\rm 4}}}
\def\etal{{\emph{et al}}}
\def\det{{\mathrm{det}}}
\def\tr{{\mathrm{tr}}}
\def\ie{{\emph{i.e.}}}
\def\aka{{\emph{aka}}}


\section{Introduction}
In a recent paper~\cite{Crab} we presented a new constraint on
Lorentz violation of order $E/M_{\rm Planck}$ in the electron
dispersion relation, based on the observation of synchrotron
radiation from the Crab nebula.  This constraint improved by a
factor of one billion the previous best constraint, which was
based on the observation of photon absorption from distant
blazars~\cite{SG,Short,Long} on the infrared background
radiation.  (The latter constraint limits the relevant Lorentz
violating parameters to be of order unity in Planck units.) Both
of these analyses were criticized in a recent paper by
Amelino-Camelia Ref.~\cite{GAC}.  The purpose of this note is to
respond to these comments, clarifying our strategy and correcting
some points that have been misconstrued in~\cite{GAC}.

\section{The synchrotron constraint}

\subsection{Theoretical Framework of Effective Field Theory}
Ref.~\cite{GAC} observes that
the synchrotron constraint of Ref. \cite{Crab}
relies on assumptions
about the {\it dynamics} of electrons and electromagnetic fields
rather than just the {\it kinematical} dispersion relations like some
other constraints.  It is stated there that if dynamical assumptions
are made then ``we are back to our starting point, we are actually
proposing a full quantum-gravity theory, with all the uncertainties
and risks of inconsistencies that plague quantum-gravity research."
It is true that constraints derived using kinematics and dynamics are
qualitatively different than constraints derived from kinematics
alone.  However, we do not agree that our assumption about low energy
dynamics is equivalent to proposing a full quantum gravity theory.

The essential assumption in Ref.~\cite{Crab} about the full quantum
gravity theory is that at
energies low compared to the Planck energy
electrons and the
electromagnetic field can be described by an effective field theory
preserving gauge invariance and rotation invariance.
The only aspect of the interactions in
this effective field theory that we use is that the modifications to
the usual Lorentz-invariant QED equations of motion are suppressed by
an inverse power of the Planck mass.

Let us elaborate a bit on how we view the significance of the
effective field theory assumption.  At low energies, physics is
well described by the standard model and general relativity,
which are both believed to be effective field theories. Since the
particle energies even in astrophysical observations are far
below the Planck energy it is reasonable to simply extend the low
energy framework that we know is accurate. Ref.~\cite{GAC}
comments that the dynamics we adopt ``appear to be consistent
with a classical and continuous spacetime, while most authors
would expect deformed kinematics at the Planck scale to be the
result of non-classical (discrete, noncommutative,...) aspects of
spacetime structure, which should have equally dramatic (but
presently unknown) consequences for dynamics." Exotic
possibilities at the Planck scale do not preclude a low energy
effective field theory description however. Some examples of this
are lattice field theories, effective field theories of the low
energy degrees of freedom of condensed matter
systems~\footnote{Ref.~\cite{GAC} states that the superluminal
dispersion relations considered in Refs.~\cite{Short,Long} are
``conceptually disfavored'' because ordinary media always lead to
subluminal velocities.  However, there exist condensed matter
systems, like Bose-Einstein condensates (see
e.g.~\cite{Liberati}) or superfluid $^3$He-A (see
e.g.~\cite{Volo}), where Lorentz invariance emerges at low
energies for quasiparticles while superluminal dispersion
characterizes the high energy propagation.}, and noncommutative
field theory~\cite{Carroll:2001ws}.  While we feel that the
effective field theory assumption is quite reasonable, it is
certainly conceivable that quantum gravity does not satisfy it.
The constraints derived in Ref.~\cite{Crab} apply only modulo the
effective field theory assumption.

The starting point for an analysis of possible quantum gravity
effects (such as Lorentz symmetry violation) using effective field
theory is to add to ordinary field theory the operators that
realize the effects.  One can then
characterize the
observational consequences. This approach has been extensively
pursued in the case of renormalizable field
theory~\cite{Kostelecky,Kbook}.

In the synchrotron case, the effective field theory being modified
is QED.  Since there are suggestions from quantum gravity that
Lorentz symmetry might be violated with a suppression by $M_{\rm
Planck}^{-1}$, we introduce in the low energy effective theory
dimension five operators that violate Lorentz symmetry.  We assume
this is the only standard symmetry that is broken, so in
particular we assume both rotation and gauge invariance are
preserved.~\footnote{ One could of course imagine that more
symmetries are broken, but it would take a conspiracy for these
to produce the usual symmetric physics when one broken symmetry
alone would not. Hence we regard it as a fruitful strategy to
begin by constraining (or looking for) minimal deviations from
standard physics .} The dispersion relations for free electrons
and photons determine most of the terms in the effective field
theory.  In addition to momentum dependence, there can be
polarization and chirality dependence of the dispersion. These
possibilities correspond to different choices of the effective
field theory. It was recently shown in~\cite{Myers-Pospelov} that
there are only three additional terms quadratic in the fields
that contribute to the dispersion relation $E^2(p)$ at cubic
order in the momentum.  In particular, the deformation parameters
for left and right circular polarized photons must be negatives
of each other, while the parameters for each electron chirality
are independent.  For the synchrotron constraint, the
polarization dependence is irrelevant in the interesting region
of parameter space.  We assumed in our synchrotron calculation
that the Lorentz violating parameters for electrons are chirality
independent. It may be possible to lift this restriction and
still get powerful constraints.

The interaction terms in the effective field theory arise from
minimal coupling (replacement of $\partial$ by $\partial +ieA$)
and non-minimal coupling in dimension five operators of the form
$\bar{\psi}\Gamma^{\a\b}\psi F_{\a\b}$, where $\Gamma^{\a\b}$ is
built from gamma matrices and possibly a factor of the preferred
timelike direction $u^\a$. All terms other than the usual minimal
coupling are suppressed by a factor of $1/M_{\rm Planck}$ hence
can be neglected for our purposes, since the unsuppressed Lorentz
invariant interaction terms dominate.

\subsection{Objections to derivation of synchrotron constraint}
There are three main theoretical objections in Ref.~\cite{GAC} to our
derivation of synchrotron radiation in the presence of Lorentz
violation.  We respond to these objections below.  First, however, we
wish to address the criticism that the observation of Crab synchrotron
emission is ``at best a promising conjecture".  We believe it is
actually much more than a conjecture.  The interpretation of the two
observed humps in the emission spectrum from Crab and other supernova
remnants as due to synchrotron and inverse Compton (IC) is the working
hypothesis in the literature (see
e.g.~\cite{AA96},~\cite{Koyama}).
This hypothesis is able to
explain the observed fluxes with success at least up to 20 TeV, with a
value of the magnetic field that is measured with consistent results
by several methods.
Uncertainties remain concerning the possible presence of
more than one population of accelerated
electrons and the specific mechanism responsible for the highest
energy IC photons (e.g. synchrotron-self-Compton (SSC) plus IC on
infrared and CMBR photons or hadronic
contributions~\cite{AA96,Tanimori}),  but the synchrotron nature of the
first hump is widely accepted.
Thus we think that our adoption of the ``standard
model'' for the Crab emission is well-justified.  We now turn to the
theoretical objections in Ref.~\cite{GAC} about the derivation of the
synchrotron constraint.

\subsubsection{Validity of heuristic formula for the
  cutoff frequency of synchrotron radiation}
Ref.~\cite{GAC} states that we assume the Lorentz invariant expression
(Eqn.~(4) in~\cite{Crab}) for the synchrotron cutoff frequency
$\o_c(E)$ is valid in the Lorentz violating case.  This is not an
assumption---it is straightforward to verify explicitly and was done.
As we state in our paper, this is a purely kinematical result.  It
involves the radius of curvature $R(E)$ of the electron trajectory,
the angular width $\delta(E)$ of the synchrotron beam, and the group
velocities of the electron and light, all of which are purely
kinematical quantities
(although the energy dependence of the first
two depends on the dynamics).

\subsubsection{Value of electron path radius of curvature $R(E)$}
%
Ref.~\cite{GAC} comments that we assume $R(E)$ is equal to its Lorentz
invariant value, and that support for this assumption is given by
proposing a new dynamics.  We are not proposing a new, arbitrary
dynamics, but merely implementing the dynamics given by the effective
field theory approach discussed above.
Since the modification of the electron field equation is strongly suppressed,
$R(E)$ remains almost unchanged. More explicitly,  gauge
invariance determines the
leading order interaction of the electrons and electromagnetic field
via minimal coupling as described above.  We show by calculation using
this minimal coupling that while $R(E)$ is not equivalent to its
Lorentz invariant value, it varies by some small relative amount which
can be neglected in computing the cutoff synchrotron frequency.

\subsubsection{Value of the opening angle $\delta(E)$ of
  the synchrotron beam}
%
Ref.~\cite{GAC} states that the relation between the opening angle
$\delta(E)$ and $E$ is assumed to be unchanged.  In fact it is not
assumed, but rather argued (albeit briefly) that the scaling
$\delta(E)\sim \gamma^{-1}(E)$ follows from the effective field
theory.  We expand on this argument here. For a fixed source term
in the electromagnetic field equation, the effective field
equation has a solution for the vector potential of the form $A =
A_{LI} + A_{dev}$, where $A_{LI}$ is the field that would be
produced by the same source using the standard Maxwell
equations.  The deviation $A_{dev} $ is a consequence of the
Lorentz violation and contains a suppression factor of $1/M_{\rm
Planck}$. Dimensional analysis indicates that it will thus be
suppressed by a factor $\o/M_{\rm Planck}$ where $\o$ is the
highest frequency in the problem.  For the synchrotron emission
from the Crab nebula, this frequency is roughly 1 GeV, leading to
a suppression of $A_{dev}$ by a factor of $10^{-19}$.  This is not
competitive with the Lorentz invariant term, hence can be
neglected. Thus the angular distribution of the radiation from a
given source will be to a very good approximation the same as it
is in the Lorentz invariant case.

In Ref.~\cite{GAC} evidence is given for large deviations of
$\delta(E)$ by viewing the synchrotron process as an off-shell
threshold phenomenon, and observing that the angular distribution
in such a process can be very sensitive to Lorentz violation.
However, it is a long way from such an observation for individual
off-shell processes to a calculation of the classical, coherent
effect of radiation from accelerating charges in a slowly varying
magnetic field.  It is clear from our field theoretic analysis
that in the end the angle sensitivity discussed in
Ref.~\cite{GAC} has no impact on the opening angle $\delta(E)$.
Unlike in threshold phenomena, there is a Lorentz invariant
zeroth order contribution that always dominates the synchrotron
emission.

\section{$\gamma$-ray absorption with the infrared background}
We turn now to the weaker constraints that can be derived from the
absorption of high energy gamma rays from blazars on the cosmic
infrared background. Ref.~\cite{GAC} argues that the type of
constraint derived in~\cite{SG,Short,Long} is conditional on
unverified assumptions about the source spectrum and IR
background. We agree that there is some uncertainty here, although
not as much as it is made out to be in~\cite{GAC}. Our reasoning
was not very explicit in~\cite{Long} however, so we elaborate here
briefly on our viewpoint, taking the opportunity to strengthen
the case somewhat.

Our starting point was the analysis carried out in
reference~\cite{SD}. There it was shown that using the most
accurate model available for the infrared background the
reconstructed spectrum of Mkn 501 shows no sign of anomalous pile
up. Moreover, it was shown recently~\cite{Kono} that the SSC model
accounts remarkably well for the intrinsic spectra of the blazars
Mkn 501 and 421 (the latter in two different states of emission)
consistently in both the synchrotron and IC regions, using the
same IR background. Hence the statement in Ref.~\cite{SG} that
there is ``no indication of Lorentz invariance breaking'' up to
20 TeV is well justified.

The constraint in question corresponds to the statement that the
soft photon threshold for absorption of a 20 TeV gamma ray should
not be shifted upwards beyond 25 meV (50 $\mu m$),~\footnote{In
Ref.~\cite{GAC} the author reported an error in our
paper~\cite{Long} saying that 25 meV photons correspond to a
wavelength of 8 $\mu m$ rather than 50 $\mu m$. This statement is
not correct.}
 the usual Lorentz invariant
threshold ($\omega_{th}=m^2/k$) for a 10 TeV photon. To justify
this statement in Ref.~\cite{Long} we said that since there is no
evidence for anomalies up to 10 TeV, it is unlikely that the
threshold for 25 meV can be raised by more than a factor of order
unity and remain consistent with the data. This was criticized in
Ref.~\cite{GAC} on the grounds that one cannot observationally
confirm absorption of any given {\it soft} photon energies, since
it is only the effect on the spectrum of {\it hard} photons that
is observed. We believe our justification of the constraint was
inadequate, but the constraint is nevertheless justified by the
absence of anomalies in the reconstructed source spectrum out to
20 TeV rather than just out to 10 TeV. (This factor of two makes a
significant difference because it appears cubed in the
constraint.)

The grounds for such a constraint follow from the shape of the IR
background spectrum reported in~\cite{SD}  If the absorption
threshold for 20 TeV gamma rays were shifted up by a factor of two
from 12.5 meV to 25 meV, that would eliminate all the absorption
from the far infrared hump of the spectrum. This would lead to a
sharp downturn in the reconstructed source spectrum above 10 TeV,
which would be inconsistent with the SSC source model. (To be more
precise about this effect it would be necessary to reconstruct
the source spectrum allowing for Lorentz violation in the
absorption on the IR background.)

\section*{Acknowledgements}
We are grateful to Giovanni Amelino-Camelia for stimulating
criticism. This work was supported in part by the NSF under grant
No. PHY98-00967.



\begin{thebibliography}{99}

\bibitem{Crab}
T.~Jacobson, S.~Liberati and D.~Mattingly,
{\em ``Lorentz violation and Crab synchrotron emission: A new constraint far  beyond the Planck scale''},
[arXiv:astro-ph/0212190].


\bibitem{SG}
F.~W.~Stecker and S.~L.~Glashow,
Astropart.\ Phys.\  {\bf 16}, 97 (2001). [arXiv:astro-ph/0102226].

\bibitem{Short}
T.~Jacobson, S.~Liberati and D.~Mattingly,
Phys.\ Rev.\ D {\bf 66}, 081302 (2002). [arXiv:hep-ph/0112207].

\bibitem{Long}
T.~Jacobson, S.~Liberati and D.~Mattingly,
{\em ``Threshold effects and Planck scale Lorentz violation: Combined  constraints from high energy astrophysics''},
[arXiv:hep-ph/0209264].

\bibitem{GAC}
G.~Amelino-Camelia,
{\em ``Improved limit on quantum-spacetime modifications of Lorentz symmetry  from observations of gamma-ray blazars''},
[arXiv:gr-qc/0212002].

\bibitem{Liberati}
C.~Barcelo, S.~Liberati and M.~Visser,
Class.\ Quant.\ Grav.\  {\bf 18}, 1137 (2001).
[arXiv:gr-qc/0011026].

\bibitem{Volo}
G.~E.~Volovik,
Phys.\ Rept.\  {\bf 351}, 195 (2001). [arXiv:gr-qc/0005091].

\bibitem{Carroll:2001ws}
S.~M.~Carroll, J.~A.~Harvey, V.~A.~Kostelecky, C.~D.~Lane and
T.~Okamoto,
Phys.\ Rev.\ Lett.\  {\bf 87}, 141601 (2001).
[arXiv:hep-th/0105082].

\bibitem{Kostelecky}
D.~Colladay and V.~A.~Kostelecky,
Phys.\ Rev.\ D {\bf 58}, 116002 (1998). [arXiv:hep-ph/9809521].

\bibitem{Kbook}
{\it Proceedings of the 2nd Meeting on CPT and Lorentz Symmetry
(CPT 01)}, Bloomington, Indiana, 15-18 Aug 2001. Ed.
A.~Kostelecky. Singapore, Singapore: World Scientific (2001).

\bibitem{Myers-Pospelov}
R.~C.~Myers and M.~Pospelov,
{\em ``Experimental challenges for quantum gravity'',}
[arXiv:hep-ph/0301124].

\bibitem{AA96}
F.A.~Aharonian, A.M.~Atoyan, Mon. Not. R. Astron. Soc. {\bf 278},
525 (1996).

\bibitem{Koyama}
K.~Koyama \etal ,
Nature {\bf 378}, 255 (1995).

\bibitem{Tanimori}
T.~Tanimori {\em et al.}, ApJ {\bf 492}, L33 (1998).

\bibitem{SD}
O.~C.~de Jager and F.~W.~Stecker,
Astrophys.\ J.\  {\bf 566}, 738 (2002) [arXiv:astro-ph/0107103].

\bibitem{Kono}
A.~K.~Konopelko, A.~Mastichiadis, J.~G.~Kirk, O.~C.~de Jager and
F.~W.~Stecker, ``Modelling the TeV gamma-ray spectra of two low
redshift AGNs: Mkn 501  and Mkn 421,'' [arXiv:astro-ph/0302049].


\end{thebibliography}
\end{document}